\documentclass[12pt,preprint]{aastex}

\slugcomment{Accepted to ApJ Letters}
\shortauthors{Howell et al.}
\shorttitle{IR Spectra of WZ Sge}

\begin{document}

\title{
Keck IR Spectroscopy of WZ Sge: Detection of Molecular Emission from the Accretion Disk
}

\author{Steve B. Howell}
\affil{WIYN Observatory and National Optical Astronomy Observatory,\\
950 N. Cherry Ave, Tucson, AZ 85719 \\ howell@noao.edu}
\author{Thomas E. Harrison}
\affil{New Mexico State University, Box 30001/MSC 4500, Las Cruces, NM 88003 \\
tharriso@nmsu.edu}
\and
\author{Paula Szkody}
\affil{Department of Astronomy, University of Washington, Box 351580, Seattle, WA 98195
\\ szkody@alicar.astro.washington.edu}


\begin{abstract}
Time-resolved IR spectroscopy of WZ Sge was obtained 
using NIRSPEC on Keck II. 
We detect CO and H$_{\rm 2}$ emission from the accretion disk placing 
WZ Sge into a rarefied class
of astronomical objects including YSOs and high luminosity early-type stars.
During the eclipse phase, the molecular emission greatly 
weakens but no firm evidence for the secondary star is seen allowing 
new limits on its luminosity to be determined.
The detection of molecular emission provides physical properties within 
the outer disk of T=3000K and N$_H$$>$10$^{10}$ cm$^{-3}$. Such a cool, dense
region, not associated with areas of H I and He I emission, provides the first 
observational confirmation of predictions made by accretion disk models.
\end{abstract}

\section{Introduction}

Infrared spectroscopy had been shown to be a powerful tool for the study
of cataclysmic variables (CV) and has opened new research avenues for our
understanding of the mass losing, secondary stars. Results to date have
revealed that these secondaries can have odd abundances,
be of very low mass, and have very cool temperatures. 

WZ Sge is a very famous, bright variable star now known to be the closest cataclysmic
variable (d=43.5$\pm$0.3 pc; Harrison et al. 2004a).
It is also the flagship TOAD (Tremendous Outburst Amplitude Dwarf novae); systems
having, among other properties, superoutbursts of 6 magnitudes or more, being highly evolved
binaries, and containing very low mass, brown dwarf-like secondary stars (Howell \&
Skidmore 2001).
WZ Sge has been observed a number of times
in the IR including a detailed time resolved spectroscopic study (Skidmore et al., 2001;
Mason et al., 2001). However, until now, its secondary star has not shown
itself directly, thereby allowing some latitude in speculation of its true nature.

Steeghs et al., (2001) observed narrow emission lines due to irradiation
of the secondary during superoutburst and found a mass ratio of
0.040 $<$ q $<$ 0.075. Patterson et al. (2001) derive a white dwarf mass of
M$_1$ = 1.0$\pm$0.2M$_{\odot}$, indicating that the secondary star in WZ Sge has 
a mass similar to those of brown dwarfs, M$_2$$\sim$0.07 M$_{\odot}$. 
Ciardi et al., (1998) used IR photometry and spectroscopy to set limits
as well which are in general agreement, albeit a bit lower ($\sim$0.05 M$_{\odot}$).

WZ Sge has the longest known inter-outburst interval of any dwarf nova.
Two leading theories to explain this behavior invoke either a very low
viscosity parameter ($\alpha$=0.001, Smak 1993; Osaki 1996), or a truncated
inner disk (Hameury et al., 1997). The latter model invokes evaporation
or a magnetosphere to truncate the inner disk, keeping it stable against
outbursts. In this configuration, the disk of WZ Sge during quiescence should
closely resemble the disks of SU UMa systems because the viscosity has
a normal value ($\alpha$=0.1). If the low-$\alpha$ model is correct, the accretion 
disks of WZ Sge-type systems are somehow different from the other SU UMa 
systems.

We have used NIRSPEC on Keck II to obtain the highest S/N, moderate
resolution K-band spectrum ever obtained for WZ Sge. We find that there
is no direct evidence in this spectrum for the detection of the secondary star
and thus we derive new limits on its luminosity. The most remarkable aspect
of these new spectra is the detection of CO and H$_{\rm 2}$ in emission 
arising from the accretion disk. This latter discovery
allows us to provide good estimates for the temperature and density
values in this previously invisible outer accretion disk region.

\section{Observations and Reduction}

We obtained twenty-eight time-resolved IR spectra for WZ Sge on 6 September 
2003 UT using NIRSPEC 
\footnote{See http://www2.keck.hawaii.edu/inst/nirspec/nirspec.html} on Keck II. 
NIRSPEC is an all-reflective, near-infrared, high-resolution spectrograph for
the Keck II Telescope designed to operate over the wavelength region
0.95 to 5.4 microns. 

We used NIRSPEC in Low Resolution (LR) mode at one grating setting 
covering the wavelength range 
from 2.0 to 2.42 microns with a slit width of 0.38 arcsec yielding 
R=2,200. The night was clear and 
appeared photometric,
as evidenced by our standard star measurements, and the seeing was near
0.6 arcsec using eyeball estimates from the slit viewer camera.
We used the standard 4 nod mode of NIRSPEC for which the star is nodded along 
the slit at four
positions from which one final spectrum is co-added and written to disk.
The scale along the slit length (spatial direction) is 0.143 arc 
seconds (4.2 \AA) per pixel.

We also obtained similar nodded spectra, directly before and after WZ Sge,
of the standard star HD 190675 to use for telluric and flux calibration. 
Flat fields, arc lamps, and
dark frames were obtained as well. Data reduction was accomplished in the usual way where we
used both the NIRSPEC instrument data reduction procedures and our own usual IR spectral
reduction methods (see Harrison et al. 2004b). 
Both resulting data sets agreed well, with the NIRSPEC reductions being
slightly noisier. We estimate our flux uncertainties to be 10-15\%.

While we had hope that each individual spectrum would be able to stand alone and that most
would reveal the secondary star in WZ Sge once and for all, we find that this is not the
case. 
Thus, we were interested in searching for any secondary star features and
opted to use summed spectra to increase the signal-to-noise.
We co-added our 28
spectra into single sums as well as four distinct orbital phase bins. 
In this section 
we discuss features visible in our co-added spectra and we 
detail their identification and 
physical nature in \S3.

\subsection{Spectral Analysis}

Figure 1 shows our total summed spectrum with no velocity correction and no smoothing.
The salient features are: a) a gentle continuum slope, b) emission lines of H I and He I,
c) emission
features of CO (Note these cannot be the Pfund series in emission see below),
and d) weak emission centered near 2.12 and 2.20 $\mu$m (We identify the latter
with H$_{\rm 2}$, the former is probably a blend of He I and H$_{\rm 2}$). 
Table 1 lists measurements of the emission line equivalent widths and 
FWHM vales 
for the lines seen in Figure 1.

To search for secondary star features, we present in Figure 2 an unsmoothed RV corrected, 
co-added spectrum from all phases. We 
used the K2 value given in Steeghs et al. (2001) to determine the
secondary's velocity at each phase. We next used the Patterson
et al. (2001) ephemeris (also used by Steeghs et al.) to figure out time vs. phase.
Note the ephemeris phase 0.0 is photometric phase 0.0, thus we
applied the -0.043 phase offset between photometric eclipse and true inferior
conjunction to our spectra.
We then Doppler-corrected each spectrum to create
one medianed (of 28 individual spectra) spectrum which we show in Figure 2.

In Fig. 2 we see that the H I and He I emission lines appear to have an odd structure, 
but this is a result of the co-added, Doppler corrections around the orbit.
We do notice that both the strong H I and He I emission lines show 
very narrow components near line center and Br$\gamma$ even has one 
sitting in its blue wing. 
Two possible weak 
absorption lines (from the secondary?) can be seen in Fig. 2 at 2.272 and 2.277 $\mu$m. 
These absorptions 
vary in strength with orbital phase, and are stronger at quadrature and during 
eclipse. However, we caution the reader as to their reality until robust
line identifications can be made, although we see similar weak 
absorption features
in the IR spectrum of EF Eri, another short period CV with a low mass, cool secondary
(Harrison et al., 2004c). 

Figure 3 shows co-added, 20\AA\ boxcar smoothed spectra for phases 0.0 and 0.5. 
The phase 0.0 spectrum is produced by only two individual spectra,
as this is all we obtained during the short eclipse phase, while the phase 0.5 result
was produced from three spectra.
We see that the velocities of the emission lines of H I  and He I match well at these
crossing phases. The phase 0.0 spectrum clearly shows evidence for a
strong decline (unmeasureable levels) in the CO emission strength 
during the eclipse. This is expected for 
lines produced in the accretion disk as they will be partially eclipsed by the secondary
during this phase. Even though it seems that the secondary star modulates the spectrum
during eclipse, no clear signature of any spectral features 
are detected during phase 0.0.
 
\section{Discussion}

\subsection{Hydrogen Emission lines}

Hydrogen emission from the accretion disk is thought to arise in a chromosphere
like structure skirting the accretion disk and/or a (dense) expanding wind
(See Warner 1995, \S2.7).
The Brackett emission lines, for example, are seen to be optically thick
as has already
been noted in Mason et al. (2001) and emanate from the hotter, denser
accretion disk locales as well as the stream-disk interaction region.
Optically thicker H I emission lines probably are produced closer to the
plane and/or in the hot spot interacting region where the temperature is
higher while the optically thin Balmer emission arises from the accretion disk
``atmosphere". The H I emission line equivalent widths listed in Table 1 compare well with
those presented in Mason et al. (2001).

\subsection{Helium Emission Lines} 

He I at 2.058 microns is strong and suggests that the 584\AA\ spectral line
is optically thick, indicative of an origin in high density regions
within the accretion disk. He I emission is generally believed to come
from the inner disk where the temperature can be high enough to produce it,
but in TOADs, the inner disk is known to be optically and materially
thin (see Mason et al., 2001, Howell et al., 1999). The hot spot
is another suggested site for the He I emission, being hot and dense, but
the double-peaked nature of the line seems to rule this out as the
sole source.

\subsection{CO Emission Bands}

CO emission is rare in astronomical objects. A few early-type, high luminosity
stars and some compact young stellar objects (YSOs) show CO in emission and share
the properties of optically thick Brackett H I emission and the presence of dust
emission. These types of objects also contain accretion disks, albeit
different in scope to those in CVs (Calvet et al., 1993).
WZ Sge is very different in most respects compared with these classes of object,
but now joins this small elite group of astronomical objects with CO emission.

The summed spectra of WZ Sge shown in Figs. 1 \& 2 
clearly show the first-overtone CO bands in emission. It is likely
that these emission features are the cause of the rising red 
continuum seen in previous low resolution K band observations of WZ Sge.
We note that the CO emission observed cannot be caused by H I Pfund
lines as the Pfund edge occurs blue-ward of the first emission band 
and blended Pfund emission will not show the band structure we observe.

CO emission is thought to occur by CO-H collisions in a relatively dense
(N$_H$$>$10$^{10}$ cm$^{-3}$), cool (T$\sim$3000-5000K) region. The site of
the emission must be separate from locations of ionized hydrogen
and helium, as
such radiation can easily destroy the CO molecule with its low
dissociation energy of 11.1 eV. Thus, the CO gas distribution
does not follow that observed for the H I and He I 
emission lines (i.e., as inferred by, say,
Doppler maps).

Modeling of CO emission in both early type, highly luminous, A and B
emission line stars and in embedded young stellar objects has shown that
the CO lines are likely to be optically thin (McGregor et al., 1988).
The He I (2.05 um)/H I Br$\gamma$ ratio in WZ Sge is 0.58, similar to
that seen in these early-type, high luminosity emission line
stars. The ratio of the eq. width of H I Br$\gamma$ to the
first overtone CO emission band is near 3 for WZ Sge (Table 1), again 
equal to that observed in the high luminosity A and B stars and in YSOs.
These facts allow us to consider similar line formation
conditions in WZ Sge compared with those modeled in the early type, 
high luminosity stars. Therefore, temperatures near, but not much greater than, 
3000 K are required to populate
these high rotational levels. Given the strength of the CO emission, these
same models show that the required vibrational temperatures are in the range
of 2000 to 5000K, in good agreement with that stated for rotational excitation.

We know in WZ Sge that the CO emission comes from the accretion disk and
is not emitted from a larger circum-binary disk. The observational evidence stems
from the fact that we see changes (dilution) in the CO emission near phase zero
(Figure 3), a time interval during which the secondary star
partially eclipses the accretion disk.

\subsection{Other Spectral Features}

Molecular H$_{\rm 2}$ emission is likely to arise from similar density and temperature
regions within the accretion disk which
produce the CO emission. We see H$_{\rm 2}$ emission in WZ Sge as the typical
low level, broad emission centered near 2.12 (He I + H$_{\rm 2}$) and 2.22 microns. 
Confirmation and further study is needed using additional observations as our
continuum S/N does not allow detailed analysis of the H$_{\rm 2}$ emission.

Observations of early-type, high luminosity stars and YSO's often show
weaker emission due to Fe II, Na I or Mg II in the near-IR. These lines
are thought to be fluorescent emission, pumped by H I Balmer continuum emission.
We do not see these weaker lines in our data (note, no Fe II lines are in the
K band), but higher S/N observations would be required.

\subsection{Physical Limits on the Secondary Star}

The distance to WZ Sge is 
43.54$\pm$0.28 pc (Harrison et al., 2004a) which yields M$_K$=11 using the
out of eclipse K magnitude of 14.0 (2MASS). To assess the contribution to the K band flux
from the emission lines, we integrated the spectral flux with and without the
H I and He I lines (using the {\it sbands} program within IRAF) and find that 
they only contribute 8\% of the flux at K. 
Using the phase 0.0 spectrum, we can derive a luminosity limit for the secondary star
using the fact that we do not see any solid spectral evidence for it during this time.

Taking the accretion
disk contribution to be $\sim$50\% of the K band flux based on a) the difference 
between our 
phase 0.0 and phase 0.5 spectra and using the fact that it is only a partial eclipse,
b) the detailed disk analysis presented in Mason et al. (2000), and c) 
the estimate of a $>$30\% 
accretion disk contribution to the K band flux as determined by 
Ciardi et al. (1998), we derive a value of
M$_K$$\ge$12 for the secondary star. This value suggests that the secondary star is 
near spectral type L6V.
Normal L6V brown dwarfs have surface temperatures near 1500K and masses of $\sim$40
M$_{Jupiter}$,
both values in accord with the theoretical predictions for WZ Sge's secondary posited by
Howell et al. (2001).

\section{Conclusions}
The new results found in this study are the identification of CO and H$_{\rm 2}$ 
emission from the accretion disk in WZ Sge and the fact that the outer parts of the disk
are as cool as 3000K during quiescence. 
The lack of detection of spectral features from the secondary star near phase 0.0,
combined with the known distance to WZ Sge, allow a rigorous luminosity limit
to be set. We find that the secondary star must be of spectral type L6V or later.
Further high S/N, K band spectroscopy of WZ Sge is needed to both confirm our
discovery and to provide detailed spectral information on the molecular
emission and for the brown dwarf-like secondary star.

IR observations of other TOADs
are needed as well to see if these same disk conditions are present. 
It is interesting to note here that the accretion disk instability model of 
Cannizzo \& Wheeler (1984) predicts accretion disk temperatures cooler
than 5000K at quiescence and our findings are the first observational
proof that such cool regions exist.
Detailed study of the accretion disk CO emission (line shapes, strengths, etc.) 
would be useful to provide 
a new physical view into the outer regions of accretion disks, regions invisible
to other spectral regimes.

\acknowledgments
Data presented herein were obtained at the W.M. Keck Observatory, which is operated
as a scientific partnership among the California Institute of Technology, the University of
California and the National Aeronautics and Space Administration. The Observatory was made
possible by the generous financial support of the W.M. Keck Foundation.
The authors want to thank R. Probst, R. Joyce, and K. Hinkle for their
IR insights and directing us to relevant results related to CO emission and R. Campbell,
G. Hill, G. Wirth for their help at Keck HQ in Waimea. 
An anonymous referee provided many good suggestions leading to a much improved paper.
The authors wish to recognize and acknowledge the very significant cultural role and
reverence that the summit of Mauna Kea has always had within the indigenous Hawaiian
community.  We are most fortunate to have the opportunity to conduct 
observations from this mountain. 

\newpage

\begin{deluxetable}{cccc}
\tablenum{1}
\tablewidth{4.5in}
\tablecaption{Measurement of IR Emission Lines\tablenotemark{a,b}}
\tablehead{
 \colhead{Line}
 & \colhead{Peak $\lambda$ (\AA)}
 & \colhead{Eq. Width (\AA)}
 & \colhead{FWHM (\AA)}
}
\startdata
\hline
He I (blue)\tablenotemark{c}  & 20531.4   & -31.3 &   65.4 \\
He I (red)\tablenotemark{c}   & 20636.1   & -31.7 &   73.6 \\
H I (blue)\tablenotemark{c}   & 21604.6   & -65.4 &   70.6 \\
H I (red)\tablenotemark{c}    & 21706.1   & -61.0 &   72.5 \\
H$_2$          & 22216.5   &  -4.4 &  115.4 \\
CO(2-$>$0)     & 22937.5   & -20.7 &  243.9 \\
CO(3-$>$1)     & 23252.3   & -20.7 &  202.9 \\
CO(4-$>$2)     & 23545.5   & -17.7 &  181.9 \\
CO(5-$>$3)     & 23859.3   & -10.6 &  147.3 \\
\hline
\enddata
\tablenotetext{a}{By convention, emission lines have negative Eq. Width values.}
\tablenotetext{b}{The Eq. Width values were measured in the phase 0.5 spectrum.}
\tablenotetext{c}{These lines have been deblended using two Gaussian components.}
\end{deluxetable}{}


\clearpage

\figcaption[]{Figure 1 - An uncorrected, unsmoothed summed spectrum of WZ Sge.
H I, He I and molecular emission due to CO and H$_{\rm 2}$ are apparent. No absorption
features are apparent. The y-axis is relative Flux in ergs/sec/cm$^{-2}$/\AA.}

\figcaption[]{Figure 2 - An RV corrected, unsmoothed summed spectrum of WZ Sge. 
We note here the typical H I and He I emission lines. Molecular emission from CO
and H$_{\rm 2}$ are seen as well. 
Two possible unidentified absorption features (not apparent in Figure 1 but seen here in
the RV corrected spectrum) are seen at 2.272 and 2.277
microns. The y-axis is relative Flux in ergs/sec/cm$^{-2}$/\AA.}

\figcaption[]{Figure 3 - Phase 0.0 and 0.5 spectra of WZ Sge. 
The phase 0.0 spectrum is the median of two and the phase 0.5 spectrum uses
three spectra allowing approximately equal comparison.
The data are smoothed by a 20\AA\ running boxcar. The y-axis is relative Flux and
the the mean levels of the two spectra are 0.0239 (0.5) and 0.0171 (0.0) with the 
phase 0.0 spectrum offset by -0.2 in flux. The CO emission greatly weakens 
during the phase 0.0 partial eclipse, indicating an origin in the accretion disk.}

\newpage

\begin{figure}
\plotone{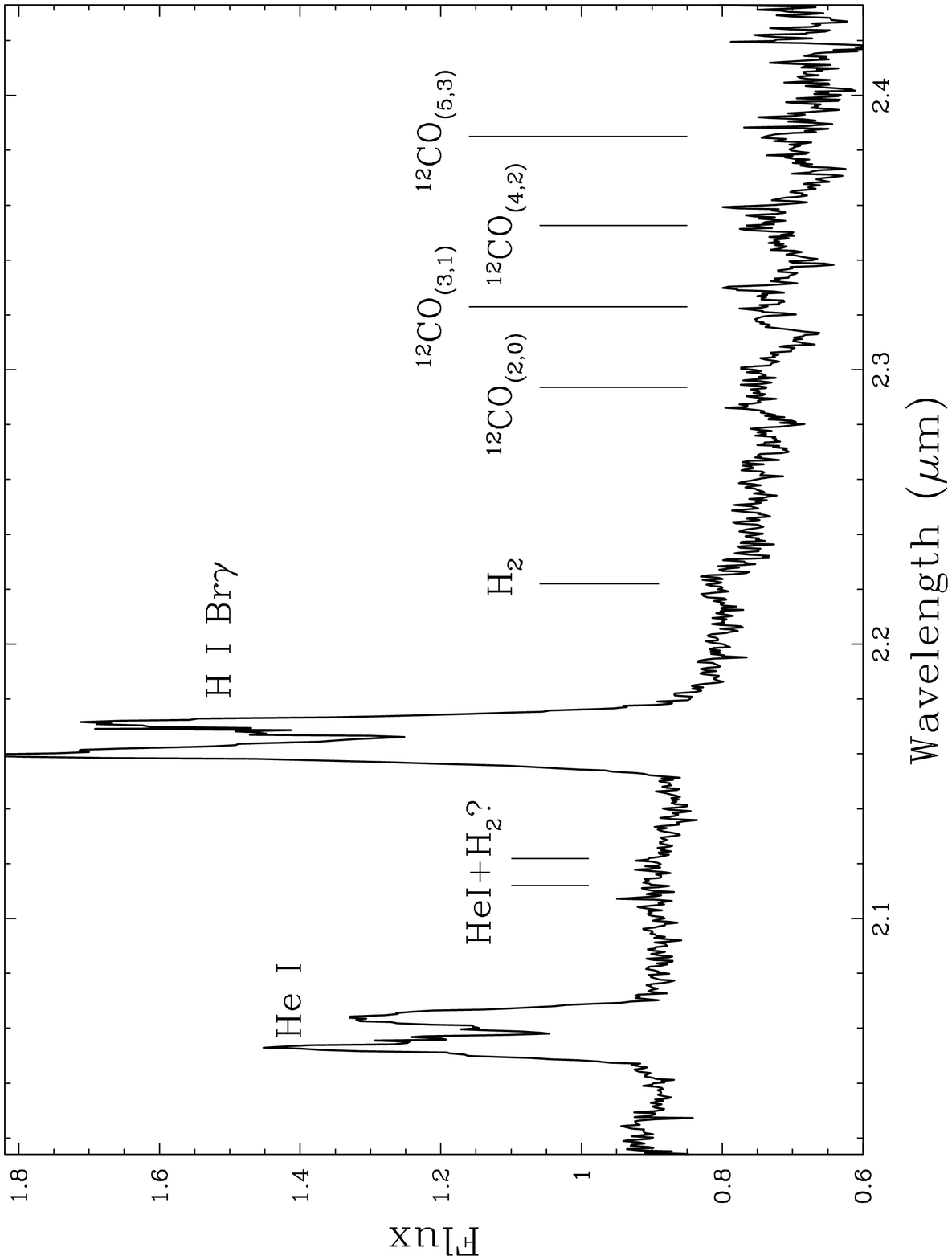}
\end{figure}

\begin{figure}
\plotone{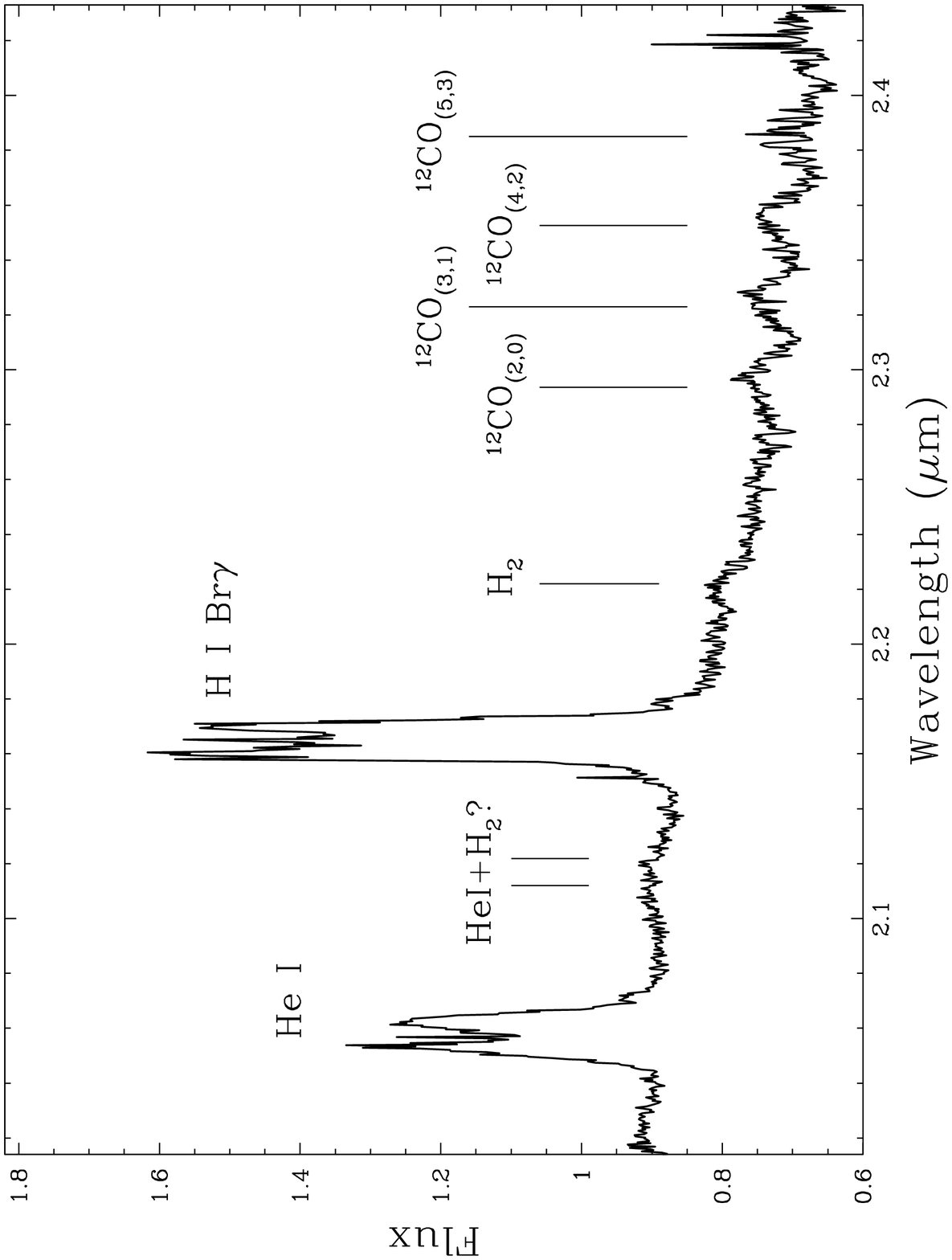}
\end{figure}

\begin{figure}
\plotone{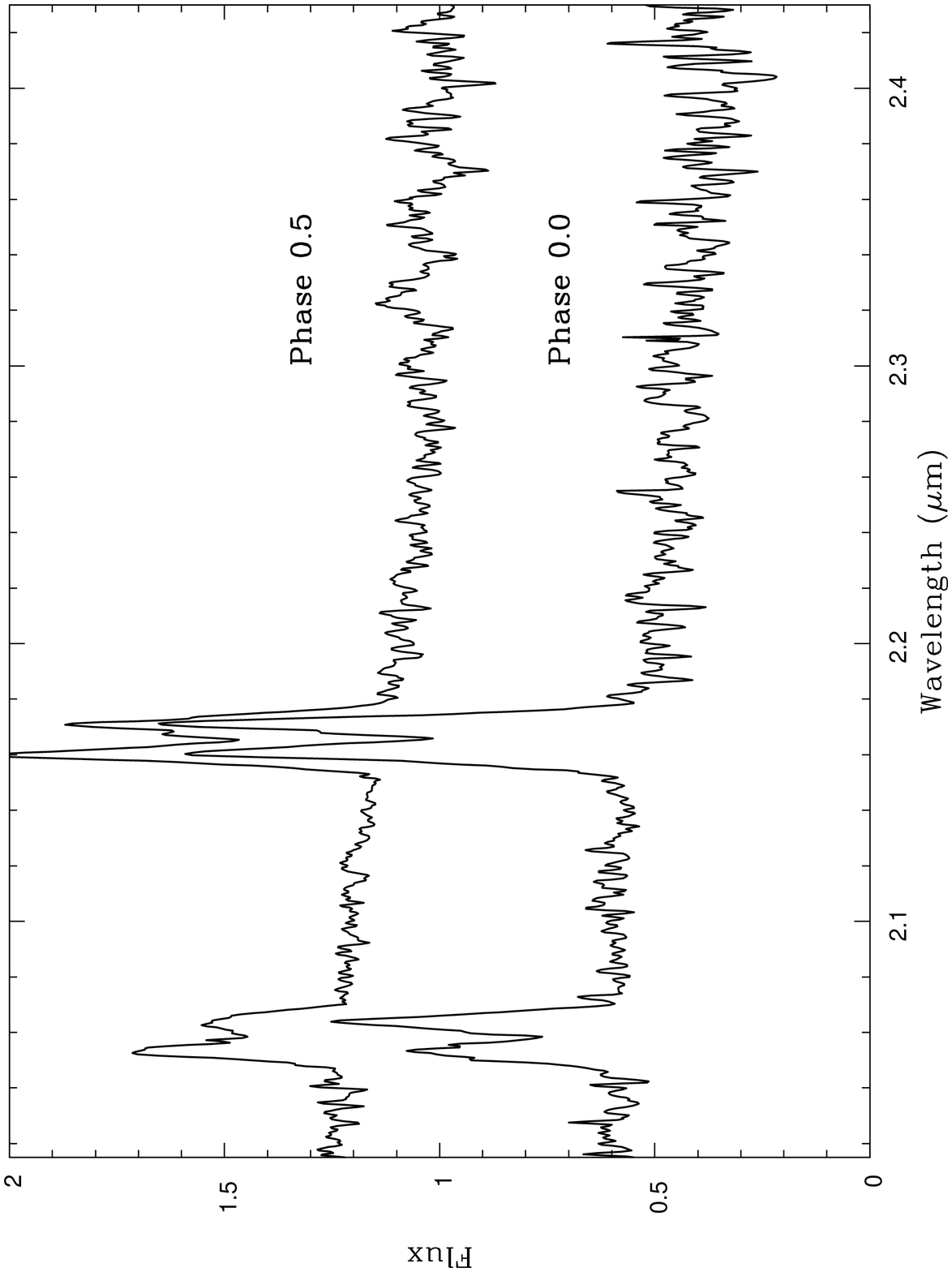}
\end{figure}

\end{document}